\def\~{{$\tilde{\phantom{a}}$}}

\documentclass [12pt] {article}
\usepackage{epsfig}
\usepackage{color}

\def\thebibliography#1{\section{References}\markboth
 {REFERENCES}{REFERENCES}\list
 {[\arabic{enumi}]}{\settowidth\labelwidth{[#1]}\leftmargin\labelwidth
 \advance\leftmargin\labelsep
 \usecounter{enumi}}
 \def\newblock{\hskip .11em plus .33em minus -.07em}
 \sloppy
 \sfcode`\.=1000\relax}
\def\upcite#1{\raise6pt\hbox{\scriptsize
\cite{#1}}}
  \def\lsim{\mathrel {\vcenter {\baselineskip 0pt \kern 0pt
    \hbox{$<$} \kern 0pt \hbox{$\sim$} }}}
    \def\gsim{\mathrel {\vcenter {\baselineskip 0pt \kern 0pt
    \hbox{$>$} \kern 0pt \hbox{$\sim$} }}}

\def\hline{\noalign{\hrule \vskip2pt}}

\def\|{\ifmmode\Vert\else \char`\|\fi}
\ifx\oldzeta\undefined                          
  \let\oldzeta=\zeta                            
  \def\zzeta{{\raise 2pt\hbox{$\oldzeta$}}}     
  \let\zeta=\zzeta                              
\fi

\ifx\oldchi\undefined                           
  \let\oldchi=\chi                              
  \def\cchi{{\raise 2pt\hbox{$\oldchi$}}}       
  \let\chi=\cchi                                
\fi

\def\frac#1#2{{#1 \over #2}}

\def\half{\ifinner {\scriptstyle {1 \over 2}}
   \else {1 \over 2} \fi}

\def\abs#1{\left\vert#1\right\vert}	

\def\simge{\mathrel{%
   \rlap{\raise 0.511ex \hbox{$>$}}{\lower 0.511ex \hbox{$\sim$}}}}
\def\simle{\mathrel{
   \rlap{\raise 0.511ex \hbox{$<$}}{\lower 0.511ex \hbox{$\sim$}}}}

\def\buildchar#1#2#3{{\null\!                   
   \mathop#1\limits^{#2}_{#3}                   
   \!\null}}                                    
\def\overcirc#1{\buildchar{#1}{\circ}{}}

\def\slashchar#1{\setbox0=\hbox{$#1$}           
   \dimen0=\wd0                                 
   \setbox1=\hbox{/} \dimen1=\wd1               
   \ifdim\dimen0>\dimen1                        
      \rlap{\hbox to \dimen0{\hfil/\hfil}}      
      #1                                        
   \else                                        
      \rlap{\hbox to \dimen1{\hfil$#1$\hfil}}   
      /                                         
   \fi}                                         %

\def\subrightarrow#1{
  \setbox0=\hbox{
    $\displaystyle\mathop{}
    \limits_{#1}$}
  \dimen0=\wd0
  \advance \dimen0 by .5em
  \mathrel{
    \mathop{\hbox to \dimen0{\rightarrowfill}}
       \limits_{#1}}}                           

\def\overlay#1#2{\ifmmode%
\setbox0=\hbox{$#1$}%
\setbox1=\hbox to\wd0{\hss$#2$\hss}\else%
\setbox0=\hbox{#1}%
\setbox1=\hbox to\wd0{\hss#2\hss}\fi%
#1\hskip-\wd0\box1 }

\def\pmb#1{\leavevmode\setbox0=\hbox{#1}%
\kern-.02em\copy0\kern-\wd0
\kern.04em\copy0\kern-\wd0
\kern-.02em\raise.04em\box0 }

\def\vereq#1#2{\lower3pt\vbox{\baselineskip1.5pt \lineskip1.5pt
\ialign{$\m@th#1\hfill##\hfil$\crcr#2\crcr\sim\crcr}}}

\def\tensor#1{\protect\@ontopof{#1}{\leftrightarrow}{1.15}\mathord{\box2}}
\def\overstar#1{\protect\@ontopof{#1}{\ast}{1.15}\mathord{\box2}}
\def\overdots#1{\protect\@ontopof{#1}{\cdots}{1.0}\mathord{\box2}}
\def\overcirc#1{\protect\@ontopof{#1}{\circ}{1.2}\mathord{\box2}}
\def\loarrow#1{\protect\@ontopof{#1}{\leftarrow}{1.15}\mathord{\box2}}
\def\roarrow#1{\protect\@ontopof{#1}{\rightarrow}{1.15}\mathord{\box2}}

\def\@ontopof#1#2#3{%
{\mathchoice
{\@@ontopof{#1}{#2}{#3}\displaystyle\scriptstyle}%
{\@@ontopof{#1}{#2}{#3}\textstyle\scriptstyle}%
{\@@ontopof{#1}{#2}{#3}\scriptstyle\scriptscriptstyle}%
{\@@ontopof{#1}{#2}{#3}\scriptscriptstyle\scriptscriptstyle}%
}%
}

\def\@@ontopof#1#2#3#4#5{%
\setbox0=\hbox{$#4#1$}%
\setbox1=\hbox{$#5#2$}%
\setbox2=\hbox{}\ht2=\ht0 \dp2=\dp0 %
\ifdim\wd0>\wd1 %
\setbox1=\hbox to\wd0{\hss\box1\hss}%
\mathord{\rlap{\raise#3\ht0\box1}\box0}%
\else   %
\setbox1=\hbox to.9\wd1{\hss\box1\hss}%
\setbox0=\hbox to\wd1{\hss$#4\relax#1$\hss}%
\mathord{\rlap{\copy0}\raise#3\ht0\box1}%
\fi
}%

\def\lambdabar{\protect\@lambdabar}
\def\@lambdabar{%
\relax
\bgroup
\def\@tempa{\hbox{\raise.73\ht0
\hbox to0pt{\kern.25\wd0\vrule width.5\wd0
height.1pt depth.1pt\hss}\box0}}%
\mathchoice{\setbox0\hbox{$\displaystyle\lambda$}\@tempa}%
{\setbox0\hbox{$\textstyle\lambda$}\@tempa}%
{\setbox0\hbox{$\scriptstyle\lambda$}\@tempa}%
{\setbox0\hbox{$\scriptscriptstyle\lambda$}\@tempa}%
\egroup
}

\def\corresponds{{\lower.2ex\hbox{=}}{\rm\kern-.75em^\triangle}}
\def\succsim{\succ\kern-.9em_\sim\kern.3em}
\def\precsim{\prec\kern-1em_\sim\kern.3em}
\def\slantfrac#1#2{\kern1em^{#1}\kern-.3em/\kern-.1em_{#2}}

\begin{document}

\begin{center}
{\Large\bf The Grating Accelerator}
\\

\medskip

Kirk T.~McDonald
\\
{\sl Joseph Henry Laboratories, Princeton University, Princeton, NJ 08544}
\\
(Sept.~14, 1984)
\end{center}

\section{Problem}

In optics, a reflective grating is a conducting surface with a ripple.
For example, consider the surface defined by 
\begin{equation}
z = a \sin{2 \pi x \over d}.
\label{p1}
\end{equation}
The typical use of such a grating involves an incident electromagnetic
wave with wave vector {\bf k} in the $x$-$z$ plane, and interference
effects lead to a discrete set of reflected waves also with wave vectors
in the $x$-$z$ plane.

Consider, instead, an incident plane electromagnetic wave with wave vector
in the $y$-$z$ plane and polarization in the $x$ direction:
\begin{equation}
{\bf E}_{\rm in} = E_0 \hat{\bf x} e^{i(k_y y - k_z z - \omega t)},
\label{p2}
\end{equation}
where $k_y > 0$ and $k_z > 0$.
Show that for small ripples ($a \ll d$), this
leads to a reflected wave as if $a = 0$, plus two surface waves that
are attenuated exponentially with $z$.  What is the relation between
the grating wavelength $d$ and the optical wavelength $\lambda$ such that
the $x$ component of the phase velocity of the surface waves is the
speed of light, $c$?

In this case, a charged particle moving with $v_x \approx c$ could
extract energy from the wave, which is the principle of the proposed
``grating accelerator'' \cite{Takeda,Mizuno,Palmer}.

\section{Solution}

The interaction between particle beams and diffraction gratings was first
considered by Smith and Purcell \cite{Smith}, who emphasized energy transfer
from the particle to free electromagnetic waves.  The excitation of
surface waves by particles near conducting structures was first discussed by
Pierce \cite{Pierce}, which led to the extensive topic of wakefields
in particle accelerators.  The presence of surface
waves in the Smith-Purcell effect was noted by di Francia \cite{Francia}.
A detailed treatment of surface waves near a diffraction grating was given
by van den Berg \cite{vandenBerg}.  Here, we construct a solution containing
surface waves by starting with only free waves, then adding surface waves
to satisfy the boundary condition at the grating surface.

If the (perfectly) conducting surface were flat, the reflected wave would be
\begin{equation}
{\bf E}_{\rm r} = - E_0 \hat{\bf x} e^{i(k_y y + k_z z - \omega t)}.
\label{s1}
\end{equation}

However, the sum ${\bf E}_{\rm in} + {\bf E}_{\rm r}$ does not satisfy
the boundary condition that ${\bf E}_{\rm total}$ must be perpendicular to
the wavy surface (\ref{p1}).  Indeed, 
\begin{equation}
\left[ {\bf E}_{\rm in} + {\bf E}_{\rm r} \right]_{\rm surface}
= 2 i E_0 \hat{\bf x} e^{i(k_y y - \omega t)} \sin k_z z
\approx 2 i a k_z E_0 \hat{\bf x} e^{i(k_y y - \omega t)} \sin k_x x,
\label{s2}
\end{equation}
where the approximation holds for $a \ll d$, and we have defined
$k_x = 2 \pi / d$.

Hence, we require additional fields near the surface to cancel that
given by (\ref{s2}).  For $z \approx 0$, these fields therefore have the
form
\begin{equation}
{\bf E} = - a k_z E_0 \hat{\bf x} e^{i(k_y y - \omega t)} \left( e^{i k_x x}
- e^{- i k_x x} \right).
\label{s3}
\end{equation}
This can be decomposed into two waves ${\bf E}_\pm$ given by
\begin{equation}
{\bf E}_\pm  = \mp a k_z E_0 \hat{\bf x} e^{i( \pm k_x x + k_y y - 
\omega t)}. 
\label{s4}
\end{equation}

Away from the surface, we suppose that the $z$ dependence of the
additional waves can be described by including a factor $e^{i k'_z z}$.
Then, the full form of the additional waves is
\begin{equation}
{\bf E}_\pm  = \mp a k_z E_0 \hat{\bf x} e^{i( \pm k_x x + k_y y 
+ k'_z z - \omega t)}. 
\label{s5}
\end{equation}
The constant $k'_z$ is determined on requiring that each of the
additional waves satisfy the wave equation,
\begin{equation}
\nabla^2 {\bf E}_\pm = {1 \over c^2} {\partial^2 {\bf E}_\pm \over
\partial t^2}.
\label{s6}
\end{equation}
This leads to the dispersion relation
\begin{equation}
k_x^2 + k_y^2 + k^{'2}_z = {\omega^2 \over c^2}.
\label{s7}
\end{equation}
The component $k_y$ of the incident wave vector can be written in terms of
the angle of incidence $\theta_{\rm in}$ and the wavelength $\lambda$ as
\begin{equation}
k_y = {2 \pi \over \lambda} \sin\theta_{\rm in}.
\label{s8}
\end{equation}
Combining (\ref{s7}) and (\ref{s8}), we have
\begin{equation}
k'_z = 2 \pi i \sqrt{ { 1 \over d^2} - \left( {\cos\theta_{\rm in} \over
\lambda} \right)^2 }.
\label{s9}
\end{equation}

For short wavelengths, $k'_z$ is real and positive, so the reflected wave
(\ref{s1})
is accompanied by two additional plane waves with direction cosines
$(k_x,k_y,k'_z)$.  But for long enough wavelengths, $k'_z$ is
imaginary, and the additional waves are exponentially attenuated in $z$.

When surface waves are present, consider the fields along the
line $y = 0$, $z = \pi/ 2 k_z$.  Here, the incident plus reflected fields
vanish (see the first form of (\ref{s2})), and the surface
waves are
\begin{equation}
{\bf E}_\pm  = \mp a k_z e^{- \pi \abs{k'_z}/2 k_z} E_0 \hat{\bf x} 
e^{i( \pm k_x x  - \omega t)}.
\label{s10}
\end{equation}
The phase velocity of these waves is
\begin{equation}
v_p = {\omega \over k_x} = {d \over \lambda} c.
\label{s11}
\end{equation}
When $d = \lambda$, the phase velocity is $c$, and $k'_z = i k_y$ according to 
(\ref{s9}).  The surface waves are then,
\begin{equation}
{\bf E}_\pm  = \mp {2 \pi a \cos\theta_{\rm in} \over d}
 e^{- (\pi/2) \tan\theta_{\rm in}} E_0 \hat{\bf x} 
e^{i( \pm k_x x  - \omega t)}.
\label{s12}
\end{equation}

A relativistic charged particle that moves in, say, the $+x$ direction
remains in phase with the wave ${\bf E}_+$, and can extract energy from
that wave for phases near $\pi$.  On average, the particle's energy is
not affected by the counterpropagating wave ${\bf E}_-$.  In principle,
significant particle acceleration can be achieved via this technique.
For a small angle of incidence, and with $a/d = 1/2 \pi$, the accelerating
field strength is equal to that of the incident wave.


\begin{thebibliography}{99}

\bibitem{Takeda}
Y.~Takeda and I.~Matsui,
{\sl Laser Linac with Grating},
Nucl.\ Instr.\ and Meth.\ {\bf 62}, 306-310 (1968).

\bibitem{Mizuno}
K.~Mizuno, S.~Ono and O.~Shimoe,
{\sl Interaction between coherent light waves and free electrons with a 
reflection grating},
Nature {\bf 253}, 180-181 (1975).

\bibitem{Palmer}
R.B.~Palmer,
{\sl A Laser-Driven Grating Linac},
Part.\ Accel.\ {\bf 11}, 81-90 (1980).

\bibitem{Smith}
S.J.~Smith and E.M.~Purcell,
{\sl Visible Light from Localized Surface Charges Moving across a Grating},
Phys.\ Rev.\ {\bf 62}, 1069 (1953).

\bibitem{Pierce}
J.R.~Pierce,
{\sl Interaction of Moving Charges with Wave Circuits},
J.\ Appl.\ Phys.\ {\bf 26}, 627-638 (1955).

\bibitem{Francia}
G.T.~di Francia,
{\sl On the Theory of some \v Cerenkovian Effects},
Nuovo Cim.\ {\bf 16}, 61-77 (1960).

\bibitem{vandenBerg}
P.M. van den Berg,
{\sl Diffraction Theory of a Reflective Grating},
Appl.\ Sci.\ Res.\ {\bf 24}, 261-293 (1971).

\end{thebibliography}
\end{document}